\documentstyle[12pt]{article}

\title{Fluctuation Effects in High Sheet Resistance Superconducting Films\footnote{To appear in the Russian review  journal {\it Uspekhi}}}
\author{James M. Valles, Jr., J. A. Chervenak,  S.-Y. Hsu\footnote{Department of Electrophysics, National Chiao Tung University, Hsinchu 300, Taiwan}, T. J. Kouh\\Department of Physics, Brown University, Providence, RI  USA}
\begin{document}
\maketitle
\begin{abstract}
 
As the normal state sheet resistance, $R_n$, of a thin film superconductor increases, its superconducting properties degrade.  For $R_n\simeq h/4e^2$ superconductivity disappears and a transition to a nonsuperconducting state occurs. We present electron tunneling and transport measurements on ultrathin, homogeneously disordered superconducting films in the vicinity of this transition.  The data provide strong evidence that fluctuations in the amplitude of the superconducting order parameter dominate the tunneling density of states and the resistive transitions in this regime.  We briefly discuss possible sources of these amplitude fluctuation effects.  We also describe how the data suggest a novel picture of the superconductor to nonsuperconductor transition in homogeneous 2D systems.   
\end{abstract}

\noindent
{\bf Introduction}

It has been known for some time that increasing the normal state sheet resistance, $R_n$, of a homogeneously disordered (as opposed to islanded) superconducting film reduces its mean field transition temperature, $T_{co}$ and superconducting energy gap, $\Delta$\cite{finkel,belitzrmp}. For $R_n$ comparable to $h/4e^2$, $T_{co}$ \cite{haviland} and $\Delta$\cite{vallesprl} approach zero and a superconductor to non-superconductor transition (SNST), commonly referred to as a superconductor to insulator transition (SIT), takes place (see Fig. 1).  Close to this transition, variations in the density of electronic states on mesoscopic length scales\cite{zhou}, disorder enhanced inelastic scattering processes\cite{belitzrmp,devereaux} and proximity to the quantum critical point separating a superconducting from a non-superconducting phase\cite{trivedi,kirkpatrick} are expected to cause large, non-mean field like fluctuations in the amplitude of the superconducting order parameter.  Finding evidence of these effects is interesting because it can provide insight into the nature of the non-superconducting phase and the physics driving the SNST. 

In this paper, we present tunneling and transport data on homogeneous films in this regime that show evidence of  large amplitude fluctuations.  The tunneling measurements indicate that the fluctuations increase the quasiparticle density relative to the cooper pair density or equivalently, increase the normal fluid fraction.  The transport measurements suggest that the paraconductance or amplitude fluctuation dominated regime of the resistive transitions grows and diverges in size at the SNST (SNST).  Taken together, the experiments intimate that amplitude fluctuation effects drive down the superfluid density, by increasing the quasiparticle density to drive the SNST in homogeneous films and the nonsuperconductor consists of weakly localized quasiparticles.  This scenario contrasts sharply with the prevalent dirty boson models of a SIT, which attribute the reduction in the superfluid density to the reduction of Josephson coupling between spatially separated film regions\cite{fisher,sitreviews}.  

\noindent
{\bf Experimental Details}

The quench condensed films that we have studied are homogeneous Bi/Sb and PbBi/Ge. They were deposited on fire-polished glass substrates held near 4 K on the cold stage of a dilution refrigerator \cite{hsuthesis,strongin}.  They consist of a two to three monolayer thick {\it in situ} deposited amorphous layer of the Sb or Ge that does not conduct at low temperatures covered by subsequent evaporations of the Bi or PbBi.  Experiments \cite{hsuthesis,strongin} indicate that these films have a uniform morphology.  First, they become electrically continuous after deposition of as little as 1-2 bulk monolayers worth of the metal as measured by a quartz microbalance.  They superconduct at 3-4 bulk monolayers.  Second, their conductance as a function of thickness indicates that the elastic mean free path is on the order of an interatomic spacing and changes very little with metal film thickness.  Finally, {\it in situ} scanning tunneling microscopy measurements on a similar system imply that the films have a smooth, rather than a granular or islanded, morphology\cite{ekinci}.   The transport and tunneling measurements that we present here were obtained using standard, four terminal, low level AC and DC techniques.  For a BCS s-wave superconductor, the normalized differential tunnel junction conductance gives a direct measure of the superconducting quasiparticle density of states, $N_s(E)=E/\sqrt{E^2-\Delta^2}$ where $E$ is measured relative to the Fermi energy and $\Delta$ is the superconducting energy gap.  At finite temperatures, the voltage dependent conductance qualitatively resembles the density of states. At low reduced temperatures, $t<.5$, the voltage at which the differential conductance crosses 1 corresponds closely to $\Delta\over e$.  

\noindent
{\bf Results}

Tunneling and transport measurements on homogeneous films strongly suggest that the transition out of the superconducting state in these films involves the collapse of the amplitude of the superconducting order parameter and a concomitant rise in the quasiparticle density.  Experiments have shown that on approaching the transition region from the superconducting side, the mean field transition temperature decreases and simultaneously, the size of the energy gap decreases\cite{dynes}.  In fact, very close to the transition, the energy gap appears to go to zero \cite{vallesprl}.  Since the energy gap gives a direct measure of the order parameter amplitude, these earlier tunneling measurements intimate that the amplitude of the superconducting order parameter goes to zero as superconductivity disappears.  

In addition, as $\Delta$ decreases, the tunneling density of states assumes an increasingly broadened BCS form\cite{hsuchervenakvalles} as shown by Fig. 2.  The differential tunnel junction conductances for these two PbBi/Ge films were obtained at a reduced temperature of $t\simeq$ 0.2 and normalized by the voltage dependent normal state conductance (see ref\cite{hsuchervenakvalles} for procedure).  The voltage dependence in the normal state results from disorder enhanced coulomb interaction effects and is well understood\cite{hsudisorder}.  Qualitatively, the tunnel junction conductances of both films exhibit the expected BCS form including a peak and a gap.  Quantitatively, the film with the larger gap fits the BCS form well.  However, the conductance of the junction on the film with the smaller energy gap, which is closer to the transition region, is higher in the gap region and the peak is substantially lower than expected for a BCS superconductor.  The inset of Fig. 2, in which the same data are plotted on a voltage scale normalized to the peak in the conductance, emphasizes these deviations. For BCS superconductors with different gaps, conductances measured at the same reduced temperature should fall upon one another when the voltage axis is scaled in this fashion.  Clearly, the peak is smaller and relatively broader and the conductance in the ``gap" is substantially higher than BCS theory predicts.  These qualitative deviations from BCS behavior continue to grow upon approaching the transition region \cite{inpreparation}.     

Fluctuation effects are also apparent in the tunneling and transport in homogeneous films at temperatures in the vicinity of $T_{co}$\cite{snspaper}.  Fig.3 shows the zero voltage bias conductance (ZBC) of a tunnel junction on and the resistance of thin homogeneous PbBi/Ge films as a function of temperature.  To facilitate comparison of the tunneling and transport data among different films, the ZBC's are normalized to their values at 8K and the resistances are normalized to their maximum value.  The ZBC is proportional to the number of quasiparticle states within 4.8k$_B$T of the Fermi energy.  In bulk conventional superconductors, the slope of the ZBC as a function of temperature changes discontinuously at $T_{co}$ where $\Delta$ appears and the ZBC decreases rapidly with decreasing temperature.  The film with the highest $T_{co}$ in Fig. 3 exhibits this behavior.  Also, the resistive transition for this film is sharp and occurs with the kink in the ZBC.  As $T_{co}$ decreases the kink in the ZBC rounds and its drop becomes more gradual.  Concomitantly the resistive transition, which occurs near the same temperature as the gap opening, becomes broader.  While some broadening near $T_{co}$ and above is expected, the broadening near the transition region is greater than conventional expectations (see below).  It is important to note that for each of the three films, the opening of the energy gap and the resistive transition midpoint, i.e. $T_{co}$, roughly coincide.  This coincidence implies that the quasiparticle density is high at and above $T_{co}$.  Correspondingly, the order parameter amplitude is very small and susceptible to thermal flutuations in the region where the broadening of the resistive transitions occurs.  In other words, this broadening of the resistive transitions corresponds to a growth of the paraconductance or amplitude fluctuation dominated portion of the superconducting transition.

Analysis of the transport data shown in Fig. 1 suggests that the size of the paraconductance fluctuation dominated regime diverges as $T_{co}$ decreases\cite{chervenak1}.  To accurately measure the size of this regime it is necessary to know the temperature dependence of the normal state conductance.  For "clean" materials this dependence is very weak and often negligible.  In these strongly disordered films the temperature dependence is too strong to ignore, but well understood.  It stems from weak localization and electron-electron interaction effects and these effects lead to a temperature dependence of the conductance that varies little with film sheet resistance.  Fig. 4 demonstrates this weak dependence, showing the conductances of a few Bi/Sb films near the transition between the superconducting and non-superconducting states.  The lowest conductance film exhibits no signs of superconductivity while the two upper films both show a conductance that increases with decreasing temperature at the lowest temperatures.  We use the temperature dependence of the conductance of the lowest conductance film shown to determine the size of the paraconductance fluctuations, $\Delta G_{sc}$, as shown in Fig. 4.   Normalized resistive transitions, obtained using $\Delta G_{sc}$ (see figure caption), are shown in Fig. 5a for films near the transition.  Clearly, the relative breadth of the superconducting transitions grows as $T_{co}$ decreases.  In the case of the film closest to the transition whose resistance drops by less than 10\% down to 0.1 K, superconducting fluctuation effects are apparent over a factor of more than 30 in temperature!  More quantitatively, the widths corresponding to the temperature interval over which the resistance drops from 90\% to 10\% of its normal state value, normalized by $T_{co}$ are plotted in Fig. 5b.  $T_{co}$ is estimated by the temperature at which the film sheet resistance has fallen to half its normal state value.  As shown, the width appears to diverge logarithmically as $T_{co}$ goes to zero. 

\noindent
{\bf Discussion}

The characteristics of the deviations of the tunneling conductances and resistively measured transitions from the simple mean-field, BCS form provide strong evidence that fluctuations in the amplitude of the superconducting order parameter grow upon approaching the SNST in homogeneous films.  Near the transition, the tunneling data cannot be fit to the simple BCS form using a single energy gap:  the peaks are too broad and there is a finite number of states in the gap region indicating states at the Fermi energy in the superconducting state.  Taken together, these imply that the amplitude of the order parameter, which is proportional to $\Delta$, is not well defined and may be zero in certain regions of a film.  The growth and apparent divergence of the width of the resistive transitions over the temperature interval where tunneling indicates that the energy gap has barely opened and hence, the order parameter amplitude is small, intimates that fluctuations in the amplitude of the superconducting order parameter become critical at the SNST.  

At the high sheet resistances considered here, $R_n\simeq R_Q$, there are a few factors that could contribute to the increasing breadth of the density of states, the broadening of the ZBC temperature dependence, and the broadening of the resistive transitions.  First, inelastic quasiparticle scattering processes become stronger with increasing sheet resistance and these processes reduce the quasiparticle lifetime\cite{devereaux}.  The reduced lifetimes broaden the structure in the quasiparticle density of states and thus, the tunnel junction conductance.  They also increase the density of low energy excitations which increases the ZBC and broaden its temperature dependence.  While such inelastic scattering effects are in the nonperturbative regime at these sheet resistances, they should still disappear at low temperatures.  Thus, they probably cannot explain all of the broadening indicated in Fig. 3 because those data were taken at a relatively low reduced temperature of .2.  

There are two other factors that could be responsible for the low temperature broadening.  Mesoscopic fluctuations in the density of electronic states near the Fermi energy could give rise to fluctuations in the amplitude of the pairing potential and the energy gap that would persist in the low temperature limit\cite{zhou}.  For $R_n\simeq R_Q$ these fluctuations are expected to be of order unity.  Also, the SNST could be a quantum phase transition with  associated critical point amplitude fluctuation effects.  Recent theories (see for example \cite{trivedi,kirkpatrick}) predict the growth of fluctuations in the superconducting order parameter amplitude that diverge at the transition.  The conductance of a large area tunnel junction measures a density of states averaged over these fluctuations.  It should be broader than the density of states for a BCS superconductor with a single gap.  The resistively measured transition would be sensitive to the accompanying spatial distribution of $T_{co}$'s in the film and would also become broader.  In two dimensions, the predicted divergence for a superconductor to normal metal transition is logarithmic\cite{kirkpatrick}, in accord with the result shown in Fig. 5b.  

The above results and discussion suggest a qualitatively different picture of the SNST for homogeneous films than the ``standard" dirty Boson models of the SIT (see for example \cite{fisher}).  In the latter, increasing the normal state sheet resistance decreases the Josephson coupling between individually superconducting islands in the film.  The concomitant reduction in the superfluid density or phase modulus leads to the growth of phase fluctuations between the islands.  If the phase fluctuations are large enough, all phase coherence is lost, Cooper pairs become localized to their individual islands, and the system becomes an insulator\cite{fisher,sitreviews}.  In contrast, in the homogeneous films considered here, the growing amplitude fluctuations increase the quasiparticle density.  These increases come at the expense of the Cooper pair or superfluid density.  At high enough sheet resistances the superfluid density is driven to zero and a nonsuperconducting phase consisting of weakly localized quasiparticles takes over. 

\noindent
\noindent
{\bf Summary}

We have presented electron tunneling and transport measurements on homogeneous films that provide evidence that fluctuations in the superconducting to order parameter amplitude grow as the amplitude collapses at the SNST.  These fluctuation effects dominate the experimentally accessible portion of the resistive transitions of the films closest to the SNST.  Theories suggest that the mesoscopic fluctuation, quasiparticle lifetime, and quantum critical fluctuation effects could each contribute to the growth of the amplitude fluctuations.  Taken as a whole, our results suggest that amplitude fluctuations become so strong near $R_n\simeq R_Q$ that they drive the superfluid density to zero to destroy the superconducting state. 

We wish to acknowledge the support of NSF grants DMR-9801983 and DMR-9502920 and helpful conversations with Fei Zhou, Brad Marston, Boris Spivak, Nandini Trivedi, Sean Ling, and Dietrich Belitz.  

\newpage
\bibliographystyle{unsrt}

\newpage

\begin{figure}
\caption{Temperature dependence of the sheet resistance of Bi/Sb films of varying thickness.  The inset shows that films with sheet resistances at 8 K much greater than 10 k$\Omega$ exhibit "insulating like" behavior and films with sheet resistances at 8 K much less than 10 k$\Omega$ superconduct with relatively sharp transitions.  The main body of the figure shows the data in the transition region between these two extremes.  The superconducting transitions move to lower temperature with increasing normal state sheet resistance, $R_n=R(8K)$.}
\label{autonum}
\end{figure}
\begin{figure}
\caption {Normalized tunnel junction conductance of superconductor(PbBi/Ge)-insulator-normal metal tunnel junctions for two different PbBi/Ge films ($T_{co}=$4.43 and 1.64 K).  The data (solid lines) were taken at a reduced temperature of $\simeq$0.2 and have been normalized by the conductance of the junction in a magnetic field high enough to suppress superconductivity.  The higher $T_{co}$ film is fit well using the BCS form for the density of states(dashed line). The inset shows the same data with the voltage axis scaled by the peak position in each data set.   If both conductances followed BCS predictions, their peaks and low bias conductances would be equal and a simple rescaling of the voltage axis would collapse the two curves together.}  
\label{autonum}
\end{figure}
\begin{figure}
\caption {Sheet resistance (lines with open circles) and tunnel junction zero voltage bias conductance (ZBC) (solid lines) as a function of temperature for 3 PbBi/Ge films.  To facilitate comparison, the sheet resistances have been normalized to their value at 8 K and the ZBC have been normalized to their value at the highest temperature at which data is shown.}
\label{autonum}
\end{figure}
\begin{figure}
\caption {Sheet conductance as a function of temperature for the three BiSb films closest to the superconducting to nonsuperconducting transition in Fig. 1 (solid lines).  The dashed lines are the conductance of the lowest conductance curve shifted to coincide at high temperatures with the conductances of the two higher conductance curves, which exhibit superconducting fluctuations.  We presume that the dashed curves provide an accurate description of the normal state conductances of each of the films and use it to calculate the superconducting fluctuation contribution, $\Delta G_{sc}$ to the temperature dependence of the conductance.}
\label{autonum}
\end{figure}
\begin{figure}
\caption {a) Normalized sheet resistance of Bi/Sb films near the superconducting to non-superconducting transition as a function of temperature.  The normalized resistance is obtained from $R/R_n=1-R\Delta G_{sc}$, where $R(T)$ is the sheet resistance and $\Delta G_{sc}$ is obtained as described in Fig. 4.  Note the increasing breadth of the transitions as $T_{co}$ decreases.  b) Normalized transition width plotted versus the temperature of the midpoint of the transition on a logarithmic scale.  The width is defined by $\Delta T=T(R/R_n=.9)-T(R/R_n=.1)$.  The dashed line is a least squares fit to the data.}
\label{autonum}
\end{figure}
\end{document}